\documentclass{revtex4}

\usepackage{amsmath}
\usepackage{graphicx} 

\begin{document}

\title{Borromean Hypergraph Formation in Dense Random Rectangles}
\author{Alexander R. Klotz }
\affiliation{Department of Physics and Astronomy, California State University, Long Beach}

\begin{abstract}
    
We develop a minimal system to study the stochastic formation of Borromean links within topologically entangled networks without requiring the use of knot invariants. Borromean linkages may form in entangled solutions of open polymer chains or in Olympic gel systems such as kinetoplast DNA, but it is challenging to investigate this due to the difficulty of computing three-body link invariants. Here, we investigate randomly oriented rectangles densely packed within a volume, and evaluate them for Hopf linking and Borromean link formation. We show that dense packings of rectangles can form Borromean triplets and larger clusters, and that in high enough density the combination of Hopf and Borromean linking can create a percolating hypergraph through the network. We present data for the percolation threshold of Borromean hypergraphs, and discuss implications for the existence of Borromean connectivity within kinetoplast DNA.

\end{abstract}

\maketitle
\section{Introduction}

The physical properties of a soft filamentous system are dictated by the degree of entanglement between its constituents \cite{tubiana2024topology}. Examples include the viscoelasticity induced by entanglements in polymer melts \cite{melts}, the tangling and untangling dynamics of living worms \cite{saad}, and the elastic curvature of kinetoplast DNA (kDNA) networks \cite{klotz}. Traditionally, entanglement between two filaments may be described the by Gauss linking number, which describes the integer number of times two closed curves pass through each other, and can be generalized to the real number of times two open curves intertwine \cite{eleni}. Certain materials share exotic entanglements that cannot be described by the linking number. For example, dense solutions of ring polymers contain interpenetrations of closed loops that affect the viscoelastic properties of the system \cite{michieletto2016topologically}, and so-called ``daisy chains'' may be formed of unlinked but deadlocked rings \cite{tom}, which cannot be separated by stretching the system. Another exotic form of entanglement is Borromean rings, which consist of three topologically connected loops in which no two share a direct topological link (Figure 1a). These three-body links cannot be detected by the Gauss linking number, which yields zero for each subset of two loops, but may be computed with more complex metrics such as the Milnor triple integral \cite{velavick}. Densely entangled polymer networks may form open Borromean rings (similar to the strands in braided hair), and a limited number of simulation studies have detected them using the Jones \cite{cao2014simulating} or HOMFLY polynomials \cite{michalke}. Recently, Ubertini and Rosa simulated dense solutions of ring polymers in which the topology may evolve through Monte Carlo operations that pass parts of chains through each other and used the three-body Jones polynomial to show that triplets of loops may form Borromean rings \cite{ubertini}. Random Borromean entanglements have not been observed experimentally. Physical realizations of a randomly linked ring polymer system may include Olympic gels of circular DNA molecules facilitated by topoisomerase II enzymes, either naturally in the  kinetoplast (the molecular chainmail network found in the mitochondria of trypanosome parasites \cite{shapiro1995structure}) or \textit{in vitro} \cite{krajina}. Dense solutions of synthetic polymers in which cyclization reactions may occur can also yield randomly formed topological links. It is known that dense packings of rings \cite{diao} or ring polymers \cite{davide2} may form a sufficient density of topological links that a single linked network may percolate across the system. This linking percolation transition has been investigated computationally in the context of kinetoplasts. It is unknown if entangled polymer solutions, kinetoplast networks or DNA-based Olympic gels contain Borromean links, and this manuscript is motivated by the possibility of their existence.

Recently, we investigated the percolation transition in Borromean networks of loops on a square lattice, in which no two loops are linked but each triplet of neighbors is \cite{donald}. Whereas Hopf-linked networks may be described as graphs in which each loop is a node and each linked connection is an edge, Borromean networks must be described as hypergraphs, in which the interdependent connections of three loops define a triangle of edges and nodes. The removal of a node or edge from a triangular hypergraph destroys the remaining edges. We found that the percolation transition occurs at a slightly higher fraction of occupied sites than for a regularly-linked network. It was subsequently proven by Bianconi and Dorogovtsev \cite{bianconi} that hypergraphs have a higher node percolation threshold than their graph counterparts in general. Our previous work exploited the regularity of the square lattice to study Borromean percolation, but or networks of randomly linked rings it is not known how likely the formation of Borromean components is, nor is it known whether Borromean networks can randomly form in the presence of regularly-linked networks. For example, if given ring is linked with every other ring in its spatial vicinity, it will be unable to share Borromean connections. It is challenging to investigate this computationally due to the complexity of computing triple link invariants and the combinatoric challenge of three-ring interactions.

\begin{figure}
    \centering
    \includegraphics[width=\textwidth]{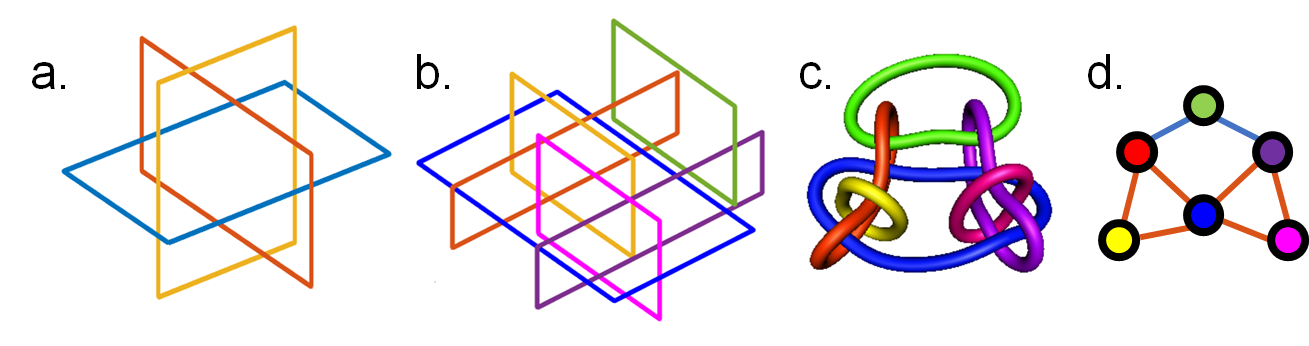}
    \caption{a. Borromean rings consisting of three perpendicular golden rectangles. b. A network of six rectangles connected by both Hopf and Borromean linking. c. The same network, with each component rendered as an elastic loop. d. Hypergraph structure of the network, in which nodes sharing a Hopf link are connected with a blue edge, and each triplet of Borromean linked rings form a red triangle of edges.}
    \label{fig:fig1}
\end{figure}

Here, we describe a minimal system with which to examine the co-existence of regular (Hopf) and Borromean linking. We study rectangles aligned with the three Cartesian planes randomly packed within a volume (Figure 1b), which can be mapped onto a system of ring polymers (Figure 1c). We can determine Borromean linking by simple comparison of the rectangles' coordinates, and construct the (hyper)graph structure of the linkages that form (Figure 1d). This allows us to examine the possibility and probability of Borromean links and percolating clusters forming in dense random networks, suggesting analogous effects for more complex polymer systems.


\section{Methods}

Squares of width 2 are initialized centered on the origin and scaled to have a given aspect ratio, defined to be greater than 1, while preserving their area. The width of the square provides a lengthscale for the system. Each rectangle then has its x, y, and z coordinates randomly permuted, such that each may lie either landscape or portrait in the XY, YZ, or ZX planes. Each rectangle is translated along each axis according to a uniform random probability distribution centered on zero with a range that determines the effective volume that the rectangles are packed into. The effective volume confines the centers of each rectangle but they may extend beyond it, particularly if their aspect ratio is large. This is repeated for $N$ on the order of several hundred or thousand to initialize a dense random network of rectangles. To examine systems analogous to the quasi-2D kinetoplast networks, we can limit the translation distribution in one of the dimensions, and translate them to fill a circle rather than a square. After initializing the rectangles, we evaluate linking and network formation through methods described below.

Consider two rectangles lying in perpendicular planes. We may define rectangle $A$ as lying in plane $ij$ with a normal along $k$, and rectangle $B$ lying in plane $jk$ with a normal along $i$. The $j$ direction is shared, $i$ and $k$ are unshared between the rectangles. We define the terms $A_{i+}$ and $A_{i-}$ as the maximum and minimum position of $A$ along the $i$ direction, and $A_{k0}$ as the location of $A$ along its normal direction. Similar terms can be defined for the other rectangle and directions. For two rectangles to have any sort of topological connection they must be colocated along their unshared axes, which is satisfied when:
\begin{align*}
     A_{i-}<B_{i0}<A_{i+} \\ \textbf{and}\ \  B_{k-}<A_{k0}<B_{k+}
\end{align*}
If the rectangles are colocated in their unshared dimension, then they are Hopf linked if one maximum or minimum of a rectangle in the shared dimension lies between the maximum and minimum of the other rectangle, satisfying either of the two following sets of conditions:
\begin{align*}
     A_{j-}<B_{j-} \ \ \&\ \  A_{j+}>B_{j-} \ \ \&\ \  A_{j+}<B_{j+} \\
     \textbf{or} \ \ \  A_{j-}>B_{j-} \ \ \&\ \ A_{j-}<B_{j+} \ \ \&\ \  A_{j+}>B_{j+}
\end{align*}
Two rectangles can be unlinked, but one can be said to \textbf{pierce} another if they lie in different planes, are colocated in their unshared dimensions, and both extrema in the shared dimension of one rectangle lie between the extrema of the other. The conditions for $A$ piercing $B$ ($A\rightarrow B$) are:
\begin{equation*}
B_{j-}<A_{j-} \ \ \& \ \ B_{j+}>A_{j+}    
\end{equation*}
Unlike linking, piercing does not commute. If the inequalities are reversed, then we may say that $B\rightarrow A$. Three rectangles $A, B, C$ with mutually perpendicular normals form a Borromean triplet if either of two conditions are met: either $A\rightarrow B\rightarrow C\rightarrow A$, or $A\rightarrow C\rightarrow B\rightarrow A$.




To efficiently evaluate linking in the dense rectangular networks, we first check each pair of rectangles for colocation, then conditionally check Hopf linking, then conditionally check colocated unlinked rectangles for piercing. Sorting rectangles by their normal axis and pre-checking each pair for colocation and linking before checking each triplet for piercing alleviates the cubic growth of the number of needed comparisons. For example, a system of 100 rectangles admits 161,700 triplets, but even in dense percolating networks only around 25 full three-rectangle piercing checks are required. Our scheme is considerably simpler than computing a knot invariant such as the Milnor triple linking integral or the Jones polynomial.


To generate a hypergraph representation of the rectangle packing, each pair of unique rectangles $A$ and $B$ is checked for Hopf linking; if two rectangles are linked then two bits, $G_{AB}$ and $G_{BA}$, on a binary $N\times N$ graph adjacency matrix $G$ are flipped. Then, each pair of unlinked normally perpendicular rectangles is checked for piercing, and if they are pierced then each rectangle $C$ that is normally perpendicular to and unlinked with the other two is checked for piercing. If the three rectangles satisfy the Borromean conditions, six bits of a binary hypergraph adjacency matrix are flipped: $H_{AB}$, $H_{AC}$, $H_{BC}$, and their transpose partners. These two matrices are used to generate a graph structure using MATLAB's `graph' function, which converts it to a table of nodes and edges that can be used to plot a graph visualization and determine the sizes of connected components using standard algorithms \cite{graph}. We track the size of the largest two components of $G$ and $H$ as a function of the number density of rectangles. The combined hypergraph is generated as $HG=2H+G$, where the magnitude of each matrix site now contains information about the weight of each edge for visualization.

\section{Results and Discussion}

\begin{figure}
    \centering
    \includegraphics[width=\textwidth]{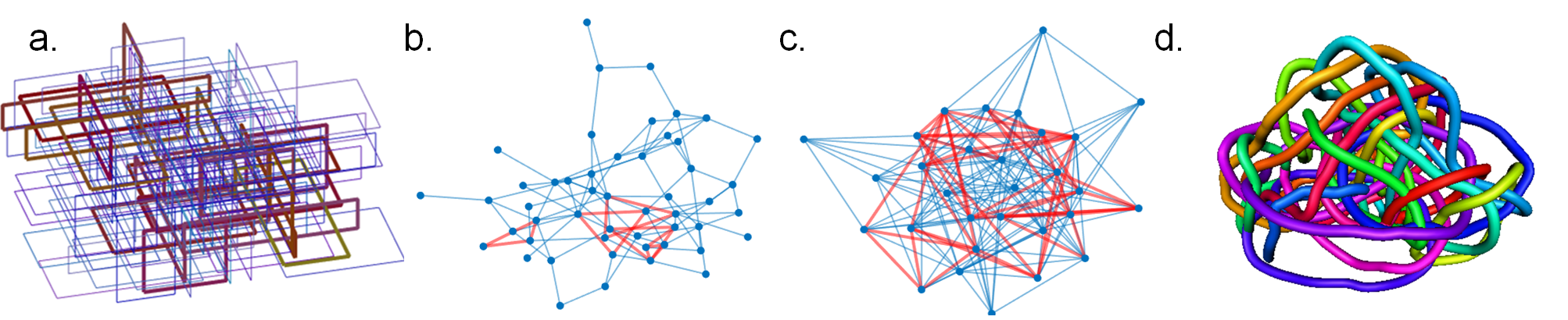}
    \caption{a. Random network of rectangles in which those participating in Borromean linking are highlighted. b. A hypergraph structure in which Hopf linked rectangles (blue edges) have formed a percolating cluster and several Borromean components (red triangles) have formed. c. A hypergraph structure in which both Hopf linked and Borromean linked rectangles have formed percolating clusters. d. Rendering of the Borromean network in c. as elastic loops.}
    \label{fig:fig2}
\end{figure}

This study was motivated by the question of whether Borromean links can form in dense randomly linked systems, and by extension whether these Borromean connections can create a percolating hypergraph. Simulations show that the answer to both these questions is ``yes.'' Figure 2a shows an example of a rectangle network, with rectangles that share Borromean connections highlighted. Figure 2b shows a hypergraph representation of a network with each node representing a rectangle, each blue edge representing a Hopf link, and each red triangle representing a Borromean triplet. In this case, the Hopf connections form a percolating cluster and a few Borromean components lie within it. Figure 2c shows a denser network in which the Borromean connections form a percolating cluster as well. The rectangles that form this percolating Borromean cluster are shown annealed and visualized as elastic filaments in Figure 2d, although it is difficult by eye to identify the connectivity.

\begin{figure}
    \centering
    \includegraphics[width=1\textwidth]{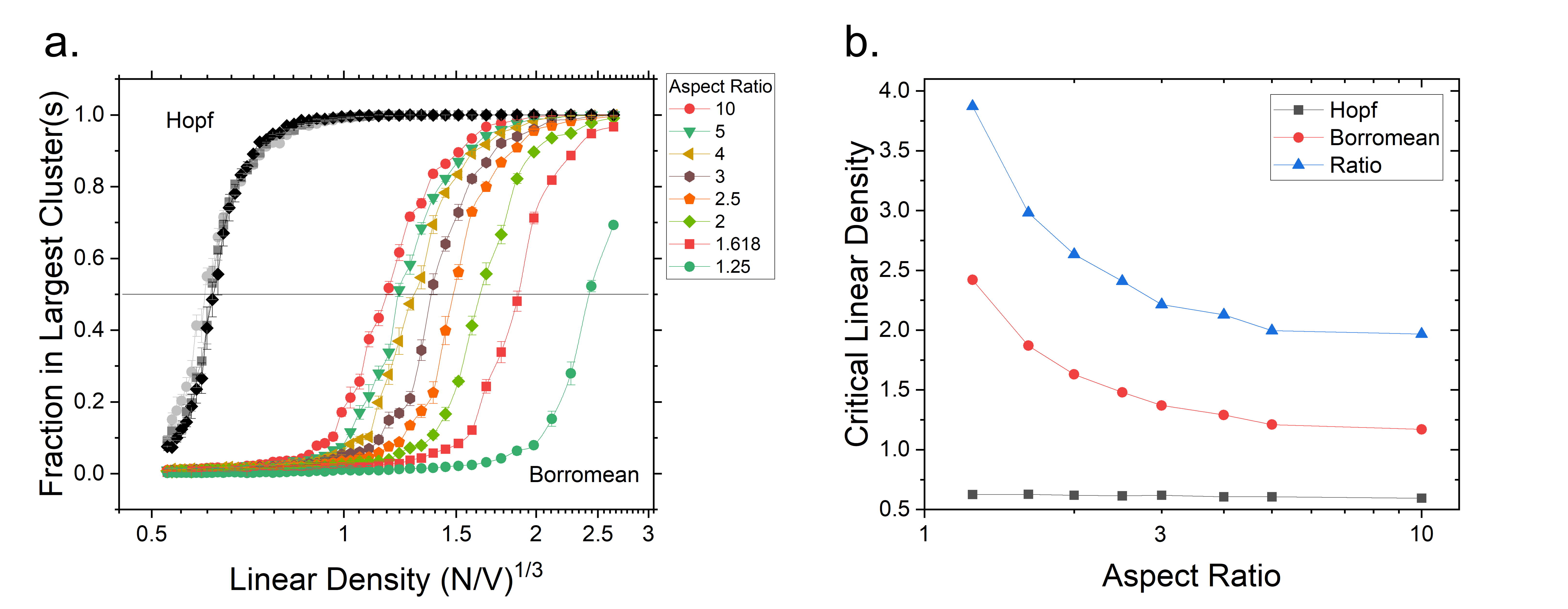}
    \caption{a. Percolation transition of rectangle dense networks as determined by the fraction of rectangles that are part of largest cluster (Hopf), or part of the largest two clusters (Borromean), as a function of the cube-root of the number density. Networks contained 500 rectangles with aspect ratios between 1.25 and 10. Three shown sets of Hopf data (1.25, 2, and 5) mostly overlap. b. Approximate linear density at percolation for Hopf and Borromean networks as a function of the rectangular aspect ratio. The Hopf link percolation threshold is largely independent of rectangle aspect ratio and the Borromean threshold is higher and dependent on aspect ratio. The dimensionless ratio between the two is also shown.}
    \label{fig:fig3}
\end{figure}

We can examine the Hopf and Borromean percolation transitions by examining the size of largest of the largest connected components as a function of the density of the rectangle packing. Although the density of two-dimensional objects in three-dimensional space is difficult to define, the number density $\rho=N/X^3$ (where X is the system scale set by the uniform distribution of translations) allows data collapse for different numbers of rings. In practice we use the cube-root of the number density for plotting, which we term the linear density. If all rectangles have a uniform size and aspect ratio, this has the effect of creating two separate but interpenetrating Borromean networks in denser systems, one in which the XY rectangles are portrait and one in which they are landscape. To track percolation in Borromean systems we examine the fraction of nodes that are in either of the two largest clusters, rather than just the largest. This is unnecessary if the rectangles are given a distribution of aspect ratios, even if the variation is limited to 10\% of the aspect ratio. Since the percolation threshold depends on the specific value or distribution of aspect ratios and we are primarily concerned with whether percolation is possible, we do not seek a detailed measure of the percolation threshold through the entire parameter space or the critical exponents at percolation. We simply use linear interpolation to find the point at which the probability of a node being in the largest component reaches 50\%.

Figure 3a shows the fraction of 500 rectangles that are found in the largest cluster or clusters, as a function of the cube-root of the number density (the cube-root simply makes visualization easier). As predicted by Bianconi and Dorogovtsev \cite{bianconi}, the Borromean networks have a higher percolation threshold. The Borromean percolation threshold may be orders of magnitude higher in density than the Hopf linking threshold, in comparison to the 1.6\% difference we observed in Borromean square lattices \cite{donald}. The critical linear densities at percolation for Hopf and Borromean networks are shown in Figure 3b. Interestingly, while the percolation threshold of Hopf linked networks appears to be independent of aspect ratio when the area is held constant, it decreases with aspect ratio for Borromean networks. There is an extremely weak but significant (-0.03 power) dependence on the Hopf critical density and this is believed to be a finite size effect. The ratio of the Borromean to Hopf thresholds appears to approach a 2-to-1 ratio in linear density, which is an 8-to-1 ratio in number density. While we do not have a rigorous explanation for the aspect independence and dependence for Hopf and Borromean percolation, we can heuristically consider that as two rectangles become longer and thinner, the probability that they overlap in the long direction increases but the probability that they overlap in the narrow direction decreases. In contrast, a lengthening of the rectangles will allow more rectangles to lie side by side within a transverse rectangle that they both pierce, increasing the likelihood of Borromean link formation with aspect ratio. If the perimeter of the rectangles is held constant as the aspect ratio is changed, rather than the area, similar trends are observed except there is a weak (0.17 power) dependence on the aspect ratio for Hopf percolation. The difference between constant-area and constant-perimeter may have to do with a subtle rescaling of the system's lengthscale when the aspect ratio is changed, affecting the definition of density. For overlapping randomly oriented rectangles in two dimensions, the percolation density also decreases with aspect ratio \cite{lin2019measurement}.

Borromean links and percolating hypergraphs were also observed in quasi-2D networks, suggesting a possibility that kinetoplast networks may contain Borromean links. To estimate the number of Borromean links contained in a kinetoplast network, should they exist, we can simulate networks with thousands of components, comparable to kDNA, and tune the number density of rectangles such that the average number Hopf links per rectangle matches the average kDNA minicircle valence, approximately 3 \cite{chen, he}. From these networks we can compute the number of rectangles participating in Borromean linking, and the total number of Borromean clusters. Figure 4a shows an example hypergraph from a quasi-2D system of 3000 rectangles and an average Hopf valence of approximately 3. Several Borromean clusters are interspersed through the network, with the largest shown in the inset. The topology of kinetoplast DNA from \textit{Crithidia fasciculata} is at this time the best-characterized, and it contains approximately 5000 minicircles. We measure the total number of links that are part of a Borromean triplet or cluster, as well as the total number of Borromean clusters. The disk-shaped networks of 5000 rectangles had an aspect ratio uniformly between 1.5 and 2.5 and an effective thickness of 0.1, and were varied in radius to change the density. As the Borromean portion of the network approaches percolation, the number of involved rectangles increases while the number of clusters reaches a maximum before approaching 1. When the mean link valence is near 3, there are roughly 1000 rectangles involved in Borromean linking, and approximately 200 Borromean clusters. This suggests that, should this system be applicable to kinetoplast DNA, the biological networks likely contain a significant number of Borromean links.

\begin{figure}
    \centering
    \includegraphics[width=1\textwidth]{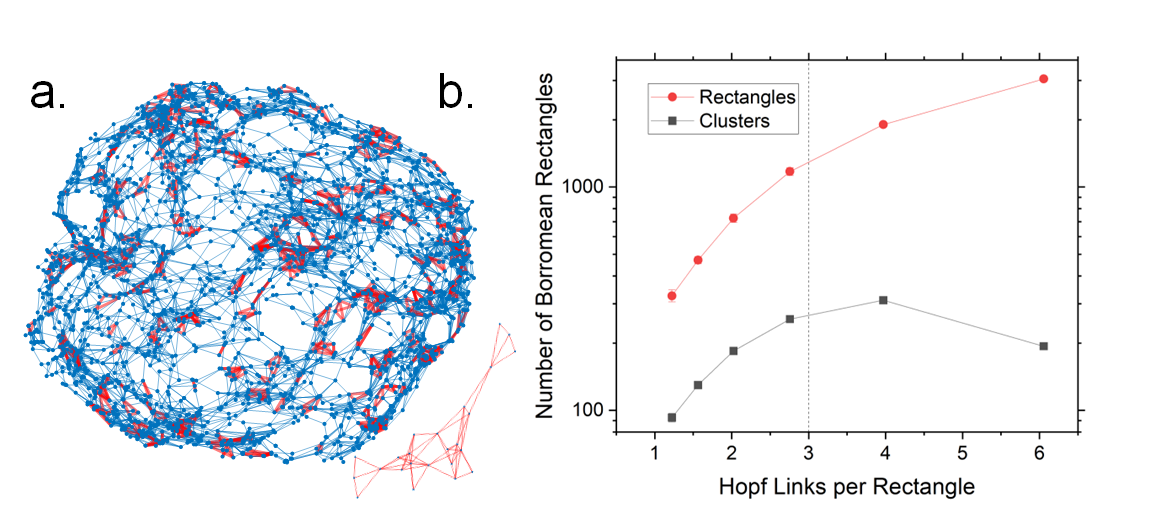}
    \caption{a. Hypergraph structure of a quasi-2D  network of 3000 rectangles with an average connectivity of three Hopf links per rectangle, with several Borromean clusters spersed throughout. The inset shows the largest Borromean cluster. b. The number of rectangles with Borromean connections, and the total number of Borromean components, as a function of the mean number of Hopf links per rectangle. The rectangles had a uniform distribution of aspect ratios between 1.5 and 2.5 in a 5000 rectangle network confined to a disk with effective height 0.1.}
    \label{fig:fig4}
\end{figure}

Finally, we may ask whether purely Borromean clusters without shared Hopf links can arise in dense networks. By definition, this would have to occur below the Hopf link percolation threshold, far below the Borromean percolation threshold. Hopf-free Borromean links can be observed in less-dense systems of rectangles, but it is a rare event even with a high aspect ratio. The largest Hopf-free Borromean cluster we observed contained 5 rectangles.

In summary, we have developed an algorithm for examining the formation of random Borromean hypergraphs in networks of dense rectangles. Our system takes advantage of the geometry of rectangles to avoid computationally difficult knot invariants. We have demonstrated the random Borromean links can form within this system, and at sufficient density will form percolating clusters within a percolating Hopf-linked cluster. This model may be extended to arbitrarily rotated rectangles, whose topological overlaps may be determined with linear algebra, and possibly ellipses for which linking and piercing may be determined based on location, normal vector, and eccentricity. Such an extension may be more applicable to ring polymer solutions. Borromean rings are the simplest Brunnian link, in which the removal of one component dissociates the entire network. Examining denser packings for 4-Brunnian, 5-Brunnian, etc. clusters may reveal higher-order hypergraphs forming at even higher densities. As computation speeds improve and the application of knot invariants to physical systems becomes more developed, it may become feasible to investigate analogous random Borromean connection and percolation in entangled polymer solutions and Olympic gels. Triple-threadings between polymer rings however may be rare compared to rectangles as the available area for piercing grows only linearly with the length of a ring polymer in a gel \cite{smrek2016minimal}. Coupled to newer quantitative experimental techniques to investigate kinetoplast DNA topology, this may lead to the discovery of naturally occurring Borromean networks.

\section{Acknowledgements}
This work was supported by the National Science Foundation, grant number 2122199. The author got the idea for this work on a Tuesday and finished the first draft on the following Saturday, and appreciates Davide Michieletto for providing feedback on it.

\bibliographystyle{unsrt}
\bibliography{rectrefs}

\end{document}